\documentclass{aa}

\usepackage{graphicx}
\usepackage{amsmath}
\usepackage{amssymb}
\usepackage{color}
\usepackage{gensymb}
\usepackage{subfigure}
\usepackage{txfonts}

\begin{document}

\title{Constraining Milky Way mass with Hypervelocity Stars}

\author{G. Fragione\inst{1}\and A. Loeb\inst{2}}

\institute{Department of Physics, Sapienza, University of Rome, P.le A. Moro 2, Roma, Italy\\ \email{giacomo.fragione@uniroma1.it}\and Astronomy Department, Harvard University, 60 Garden St., Cambridge, MA 02138, USA\\ \email{aloeb@cfa.harvard.edu}}

\date{Received}

\abstract
{Although a variety of techniques have been employed for determining the Milky Way dark matter halo mass distribution, the range of allowed masses spans both light and heavy values. Knowing the precise mass of our Galaxy is important for placing the Milky Way in a cosmological $\Lambda$CDM context.}
{We show that hypervelocity stars (HVSs) ejected from the center of the Milky Way galaxy can be used to constrain the mass of its dark matter halo.}
{We use the asymmetry in the radial velocity distribution of halo stars due to escaping HVSs, which depends on the halo potential (escape speed) as long as the round trip orbital time is shorter than the stellar lifetime, to discriminate between different models for the Milky Way gravitational potential.}
{Adopting a characteristic HVS travel time of $330$ Myr, which corresponds to the average mass of main sequence HVSs, we find that current data favors a mass for the Milky Way in the range $(1.2$-$1.9)\times 10^{12} \mathrm{M}_\odot$.}
{}


\maketitle

\section{Introduction}

Hypervelocity Stars (HVSs) are defined as stars able to escape the gravitational well of the Milky Way (MW). Theoretically predicted by \citet{hil88} as the consequence of interactions of binary stars with the massive Black Hole (BH) in the Galactic Centre (GC), HVSs were first observed by \citet{brw05}. More then $20$ HVSs at distances between $50$ and $120$ kpc from the GC, and velocities up to $\approx 700$ km s$^{-1}$, have been found (Multiple Mirror Telescope (MMT) spectroscopic survey \citep{brw10,brw14}). A similar number of bound HVSs, i.e. stars ejected by the same mechanism of unbound stars, but with velocities below the Galactic escape speed, have been observed \citep{brw07a,brw07b,brw14}. The MMT targets stars with the magnitudes and colors of $2.5-4$ M$_{\odot}$ late B-type stars, since they should not exist at faint magnitudes in the outer halo far from star-forming regions unless they were ejected to that location. Given the MMT target selection, the sample stars could be either Main Sequence (MS) B stars, evolved Blue Horizontal Branch (BHB) stars or blue stragglers, while only a few of them have a defined stellar type. Recent studies have started to investigate low-mass HVS candidates \citep{pal14,li15,zie15}. The physical mechanisms responsible for the production of the observed HVSs are still debated. However, due to their extreme velocities, the origin of HVSs involves strong dynamical interactions probably with a single or binary BHs in the GC \citep*{yut03,gil06,gil07,oll08,sar09,gin12,cap15,fra16,frg16,fck16} or in a nearby galaxy \citep*{she08,bou16}.

The study of HVSs can provide clues about the process responsible for their production in the Galactic Centre region \citep*{gou03}. As their orbits are completely determined by the MW potential, \citet{gne05} and \citet{yum07} suggested to use the kinematics of HVSs to probe the Galactic potential triaxiality. Moreover, HVSs can be used to constrain the mass distribution of our Galaxy, which is still highly uncertain. \citet{gne10} used the MMT sample to constrain the Galactic mass profile out to $80$ kpc. In this paper we apply the method proposed by \citet{per09} to discriminate among different Galactic potential models, with a focus on measuring the total dark halo mass.

A variety of techniques have been employed for determining the MW dark matter halo mass distribution, but, nevertheless, the range of allowed values spans both light ($\lesssim 10^{12}$ M$_{\odot}$) and heavy ($\gtrsim 2 \times 10^{12}$ M$_{\odot}$) values. Several classes of kinematic tracers, such as tidal streams and globular clusters, have been used for this purpose, but suffer from systematics caused by the lack of reliable tangential velocity and distance measurements \citep*{gib14}. Knowing the precise mass of our Galaxy is important for placing the MW in a cosmological $\Lambda$CDM context. Although the difference in mass between light and heavy halo masses is just by a factor of $2-3$, such a factor leads to a major difference in the efficiency conversion of baryons into stars (higher for lighter haloes), places the Large Magellanic Cloud and the Leo I dwarf spheroidal on unbound (light halo) or bound (heavy halo) orbits and can or cannot solve the Too-Big-to-Fail problem \citep*{tay15}. This latter problem is one of the two prominent challenges concerning the satellite galaxies in the MW and consists in the fact that the most massive subhaloes of numerical simulations, which in typical galaxy formation models would host the most luminous satellites, are too dense to be dynamically consistent with observations of any of the known MW companions \citep*{boy11}. The other challenge is the so-called Missing Satellite Problem, as $\Lambda$CDM model predicts hundred of subhaloes but a smaller set of galaxies is observed \citep*{kly99}. Several possible solutions have been suggested, including uncertainties in the mass of the MW halo \citep*{guo15,kan16}.

In this paper, we study the kinematics of HVSs in the Galaxy as a probe of the MW halo mass. The outline of the paper is as follows. In Section 2, we describe the method we use to discriminate among different halo masses. In Section 3 we present our models for the MW gravitational potential. In Section 4, we perform numerical simulations of HVSs motion to study their kinematics and provide results. Finally, in Section 5, we summarize our main conclusions.

\section{Method}

Theoretical calculations by \citet{yut03} suggest that the HVSs ejection rate, both in case the source at the GC is a single or a binary BH, is $\approx 10^{-5}-10^{-4}$ yr$^{-1}$. If HVSs ejections are continuous and isotropic, their number density is \citep{brw15}
\begin{equation}
n(r)=\frac{dN/dt}{4\pi r^2 dr/dt}\approx \frac{8}{(r/{\rm kpc})^2}\ {\mathrm{kpc}}^{-3},
\end{equation}
implying that HVSs are rare objects. The MMT survey \citep{brw14} targeted stars luminous enough to be observed in the Galactic halo where the relative number of HVSs is expected to be higher. \citet{dea14} found evidence for a very steep outer halo density profile, implying that the relative frequency of HVSs is much higher in the outer halo than in the inner halo. Moreover, at Galactic latitudes $|b|\gtrsim 30^{\circ}$, the survey is less likely to be contaminated by the disk and runaway stars \citep{brm09,ken14}.

We assume that HVSs are ejected from the GC \citep{brm06}, and escape the MW if the ejection velocity $v_{ej}$ is higher than the local escape speed $v_{esc}(r)$, which depends on the Galactic potential. While unbound stars will leave the MW, bound stars will reach the apocentre of their orbit and then return back to the GC with a negative radial velocity \citep{brm09,ken08,brw15}. The observations reveal a significant asymmetry in the tail of the velocity distribution of the sample stars. In particular, there is a significant lack of stars with $v_r< -275$ km s$^{-1}$ in Galactocentric coordinates \citep{brw07a,brw07b,brw10}. We divide the stars of the sample to outgoing stars with positive radial velocities in Galactocentric coordinates, and ingoing stars with negative radial velocities. \citet{per09} proposed a method that uses the observed asymmetry between ingoing and outgoing stars to discriminate among different Galactic potential models. Such asymmetry originates both from the MW gravitational potential as well as from the finite lifetime of HVSs.

Bound stars can be spotted either as outgoing stars or ingoing stars, according to when they are observed in their orbit \citep{ken14}. However, unbound HVSs can be observed only as outgoing. Therefore, an asymmetry in the distribution of ingoing and outgoing stars is expected if HVSs are continuously ejected from the GC \citep{per09}. Whereas bound stars are expected to be symmetrically divided between ingoing and outgoing stars, unbound HVSs can have only positive velocities. Furthermore, while unbound HVSs are not limited in ejection velocity, except for the limitations of the assumed ejection model, the bound stars must satisfy $v_{ej}<v_{esc}(r)$. At a given Galactocentric distance $r$, ingoing stars can be observed with a negative velocity whose amplitude is at maximum $v_{esc}(r)$, which depends on the Galactic gravitational potential. As consequence, for a given MW model, no ingoing stars are expected to be found below the curve $-\nu(r)=-v_{esc}(r)$ in the v-r plane. Therefore, the asymmetric distance-velocity distribution can be used to directly constrain the Galactic potential \citep{per09}.

However, some stars may disappear from view because they evolve to a different stellar type \citep{ken08,brm09}. For example, the finite lifetime of MS stars $t_{*}\propto m^{-\alpha}$, with $\alpha\approx 3$, implies that massive stars ejected from the GC can not reach large Galactocentric distances and fall back toward the GC before leaving the MS \citep{brm06}. The MMT targeted stars that could be late-type MS B stars with masses in the range $2.5-4$ M$_{\odot}$, for which the maximum travel time would be $t_{*}\approx 1-6\cdot 10^8$ yr \citep{brw10,brw14}. The asymmetry in the velocity-distance distribution is still expected but the cutoff $-\nu(r)$ will also depend on the finite travel time. Moreover, stars of different types have different travel times and will lead to distinct distance-velocity cutoffs, providing independent probes of the Galactic potential \citep{per09}.

In conclusion, different Galactic potential models give different $v_{esc}(r)$, which, combined together with different travel times, lead to peculiar cutoffs in the v-r plane. In this paper we apply the method proposed by \citet{per09} to current data on halo stars, with a focus on measuring the dark halo mass. We draw critical lines for HVSs both as function of the dark halo mass $M_{DM}$ and of the stars travel time $t_*$. Whereas \citet{per09} look for the best fit model that shows the largest asymmetry, we consider the one that gives compatible asymmetric distribution of stars $\Delta$ and number of high-velocity outliers $\Gamma$ in the MMT sample (see Section 4 for details) to constrain the MW mass.

\section{Models for the MW gravitational potential}

As described in the previous section, HVSs data can be used to constrain the MW potential. We describe the MW potential with a 4-component model $\Phi(r)=\Phi_{BH}+\Phi_{bul}(r)+\Phi_{disk}(r)+\Phi_{NFW}(r)$ \citep{ken08,ken14}, including the potential of the central BH
\begin{equation}
\Phi_{BH}(r)=-\frac{GM_{BH}}{r},
\end{equation}
where $M_{BH}=4 \times 10^6$ M$_{\odot}$, the bulge
\begin{equation}
\Phi_{bul}(r)=-\frac{GM_{bul}}{r+a},
\end{equation}
where $M_{bul}=3.76\times 10^9$ M$_{\odot}$ and $a=0.1$ kpc, the disk
\begin{equation}
\Phi_{disk}(R,z)=-\frac{GM_{disk}}{\sqrt(R^2+(b+\sqrt{c^2+z^2})^2)},
\end{equation}
where $M_{disk}=5.36\times 10^{10}$ M$_{\odot}$, $b=2.75$ kpc and $c=0.3$ kpc, and the dark matter halo
\begin{equation}
\Phi_{NFW}(r)=-\frac{GM_{DM}\ln(1+r/r_s)}{r}.
\end{equation}

While the parameters of the baryonic components (BH+Bulge+Disk) of the potential are kept fixed, the mass parameter (not to be confused with the total halo mass) of the Navarro-Frenk-White (NFW) dark halo $M_{DM}$ \citep*{nav97} is varied in the range $(0.6$-$1.8)\times 10^{12} \mathrm{M}_\odot$. The scale radius $r_s$ changes accordingly so that the Galactic circular velocity at Sun's distance ($8.15$ kpc) would be $235$ km s$^{-1}$ \citep{rei14}, while the virial radius $r_{200}$ is defined as the radius within which the enclosed average density is $200$ times the mean matter density in the Universe. The concentration is defined as $C=r_{200}/r_s$; see Table 1 for the cases under consideration.

\begin{table}
\caption{Parameters of the MW's NFW Dark Halo}
\centering
\begin{tabular}{c c c c}
\hline\hline
$M_{DM} (10^{12}$ M$_{\odot})$ & $r_s ($kpc$)$ & $r_{200} ($kpc$)$ & $C$\\
\hline
0.6  & 14.4 & 221.2 & 17.4\\
0.8  & 17.4 & 233.8 & 13.5\\
1.0  & 20.0 & 248.6 & 12.4\\
1.2  & 22.4 & 261.3 & 11.7\\
1.4  & 24.6 & 272.6 & 11.1\\
1.6  & 26.6 & 282.8 & 10.6\\
1.8  & 28.5 & 292.0 & 10.2\\
\hline
\end{tabular}
\label{tab1}
\end{table}

\section{Results}
\label{sec:res}

\begin{figure}
\centering
\subfigure{\includegraphics[scale=0.68]{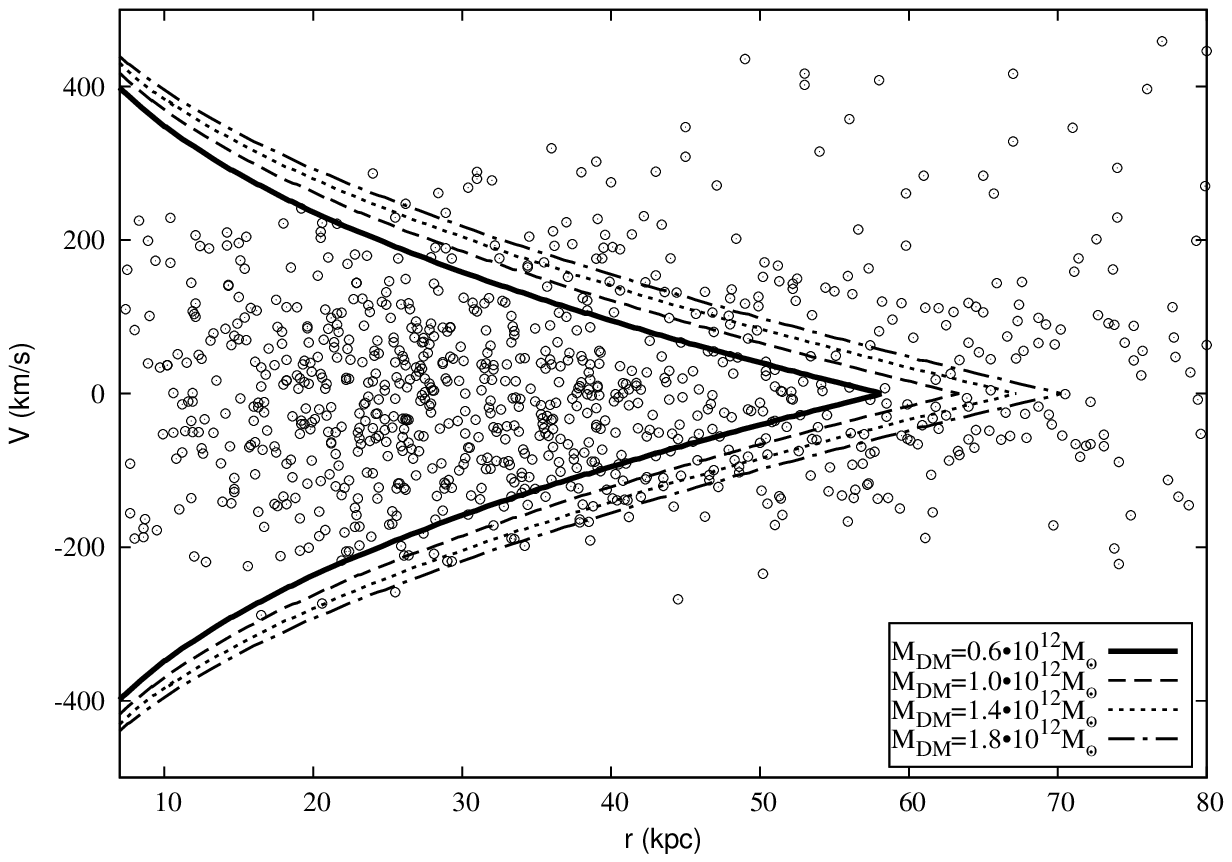}}
\subfigure{\includegraphics[scale=0.68]{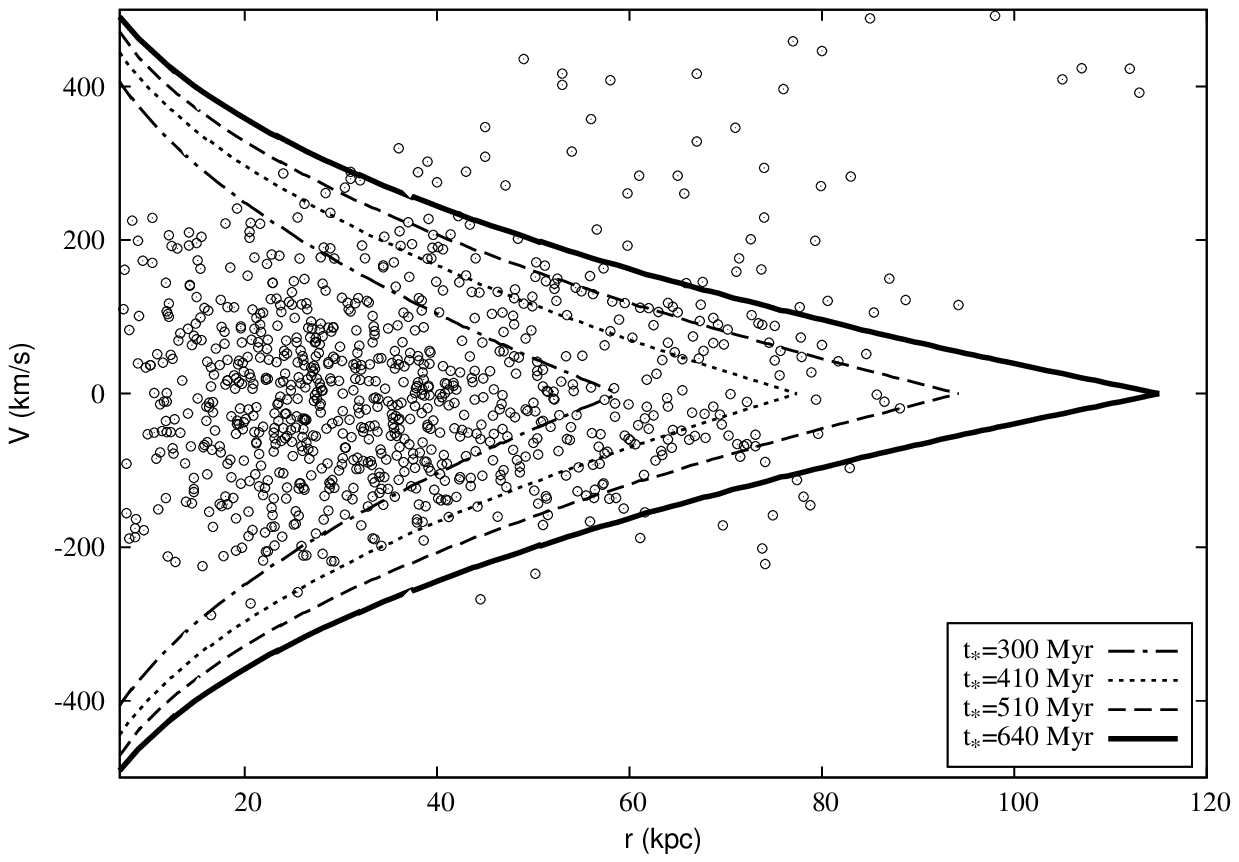}}
\caption{Critical lines for different halo masses (top) and star travel times (bottom). In the top panel, the HVSs travel time is fixed to $t_*=330$ Myr, while, in the bottom panel, the mass parameter  of the NFW dark halo $M_{DM}$ is set to $10^{12}$ M$_{\odot}$. Circles are observed distances and velocities of the MMT star sample \citep{brw10,brw14}.}
\end{figure}

We use the publicly available MMT data (Table 1 of \citet{brw10}). The MMT stars could be either MS B stars, evolved BHB stars or blue stragglers \citep{brw14}. The \citet{brw10} data presents photometric and kinematic measurements of halo stars, with positions estimated both for the case the stars are BHB or blue stragglers, and the likelihood $0<f_{BHB}<1$ of each star being BHB. We take the BHB estimated distance if $f_{BHB}>0.5$, and the blue straggler distance if $f_{BHB}\le 0.5$. Moreover, we remove the sub-sample of stars that have an estimated Galactic velocity $v>275$ km s$^{-1}$. Finally, we update the distances and velocities of this sub-population with the newer data in Table 1 of \citet{brw14}.

It is important to note that stellar type plays a crucial role since the distances can be inferred only once the luminosities of the observed stars are known. To estimate luminosities and masses, the fitted stellar atmosphere parameters need to be compared to theoretical star tracks. Such a comparison can be done once the stellar type is assumed. As discussed, the cutoff in the distance-velocity distribution depends not only on the Galactic potential, but also the travel time, which depends on the stellar type. For MS stars, $t_*$ is directly related to the star mass $m_{*}$ by $t_{*}\propto m_*^{-\alpha}$, with $\alpha\approx 3$. However, only few stars of the sample have a defined stellar type. For example, HVSs in the MMT survey are probably all MS B stars based on stellar rotation \citep{brw14}. Future identification of HVSs types will improve the constraint on the halo mass. Since $-\nu(r)$ has a two parameter dependence, we have to fix either the propagation time $t_*$ or $M_{DM}$ in order to draw critical lines in the v-r plane.

Figure 1 shows the critical lines for different halo masses and propagation times, along with the distribution in the v-r plane of the stars in the MMT sample. The bottom arc in Fig.1 corresponds to the critical line $-\nu(r)$, whereas the top arc to $\nu(r)$. The critical lines show a clear dependence on $M_{DM}$ and $t_*$. The top figure shows the dependence on the halo mass, if $t_*=330$ Myr for all the stars. The zero-velocity point, i.e. the distance at which $-\nu(r)$ crosses the x-axis, is an increasing function of the halo mass. In order to interpret these results, let us consider two different halo masses $M_{DM,1}>M_{DM,2}$ and a bound HVS. Once the baryonic content is fixed, $v_{esc}\propto M_{DM}^{1/2}$. As consequence, $v_{es,1}>v_{es,2}$ and bound HVSs can be produced in the GC with a higher ejection velocity, since they have to satisfy the constraint $v_{ej}<v_{esc}(r)$. Since $t_{*}$ is fixed by the stellar type, stars ejected in a Galaxy with $M_{DM,1}$ are able to reach farther and fall back toward the GC in a shorter time, given their possible higher ejection velocities. Given that, the zero-velocity point is located at larger Galactocentric distances for heavier haloes. On the other hand, the shape of the critical lines depends also on the maximum HVS travel time. The bottom panel of Fig.1 shows the dependence on $t_*$, if the halo mass is fixed to $10^{12}$ M$_{\odot}$. The zero-velocity point is an increasing function of the propagation time since stars are able to reach farther in their Galactic orbits within a longer $t_*$, before returning back to the GC with negative radial velocity. As discussed in the previous section, for MS stars the propagation time is directly related to the mass $m_*\propto t_*^{-\beta}$, with $\beta\approx 1/3$. Hence low-mass MS stars are able to reach larger distances in the Galactic halo within the MS lifetime, before returning back to the GC with negative radial velocity, compared to massive MS stars.

\begin{figure}
\centering
\vspace{-0.5cm}
\subfigure{\includegraphics[width=8.75cm,height=7.5cm]{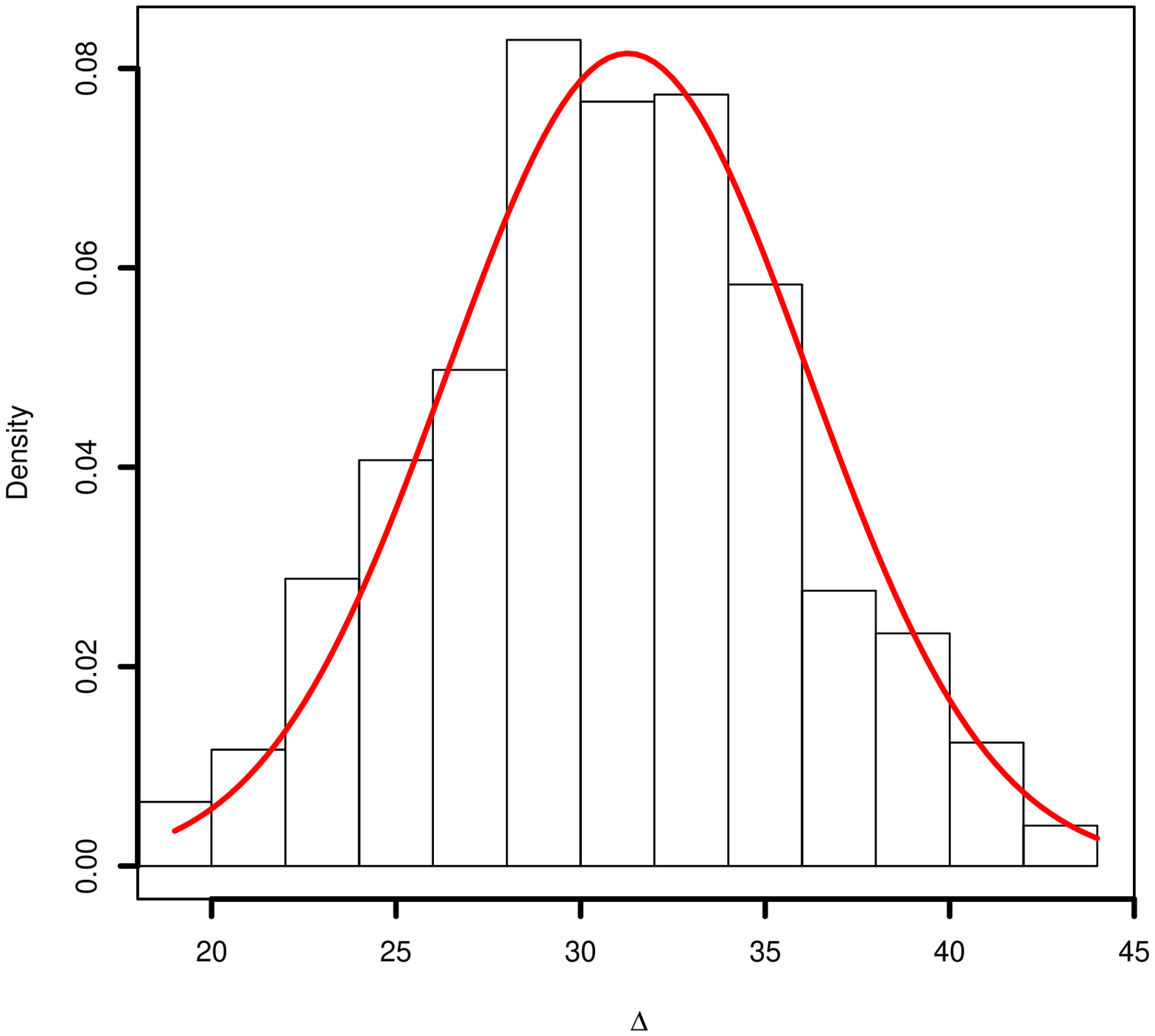}}
\subfigure{\includegraphics[width=8.75cm,height=7.5cm]{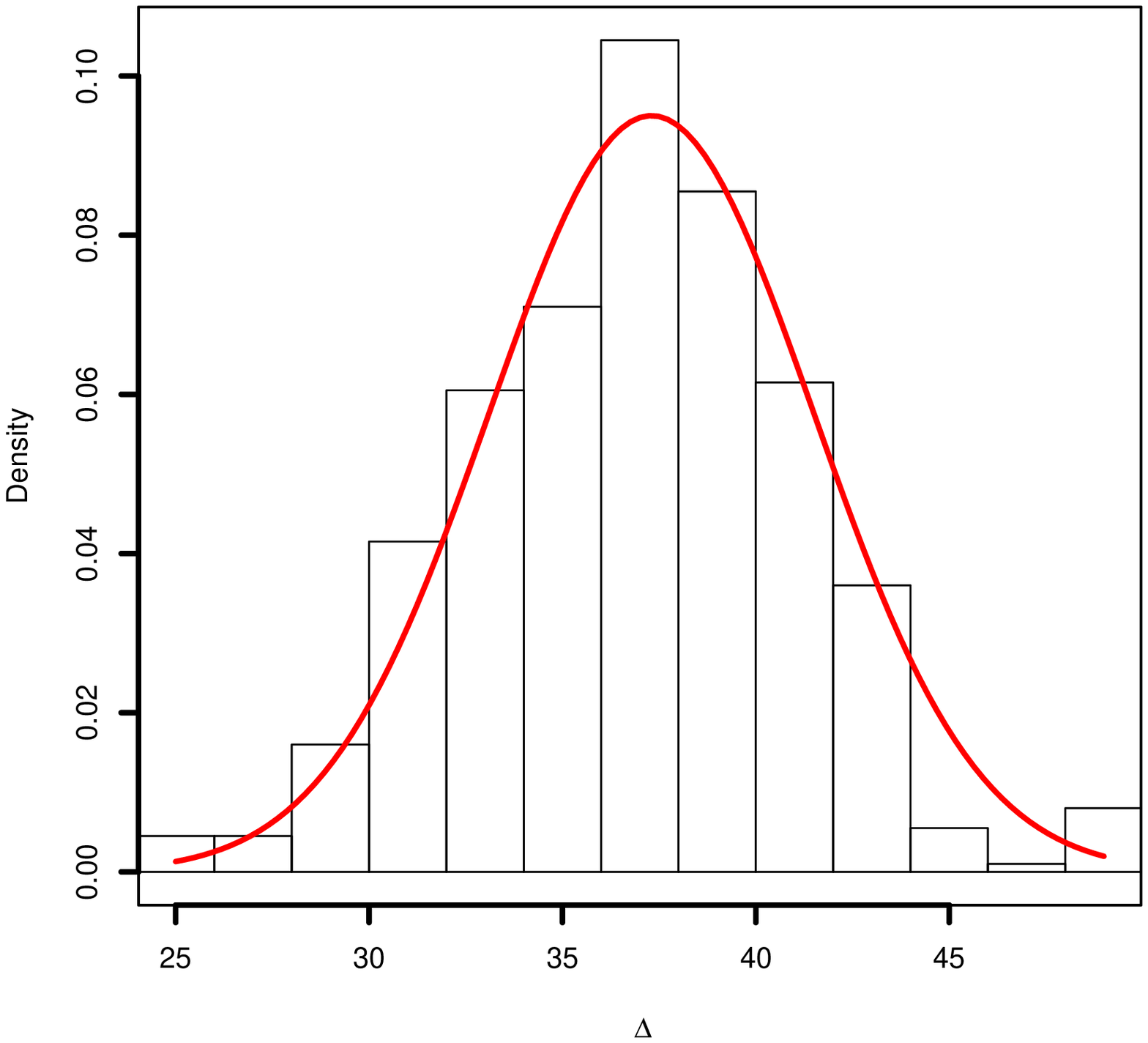}}
\subfigure{\includegraphics[width=8.75cm,height=7.5cm]{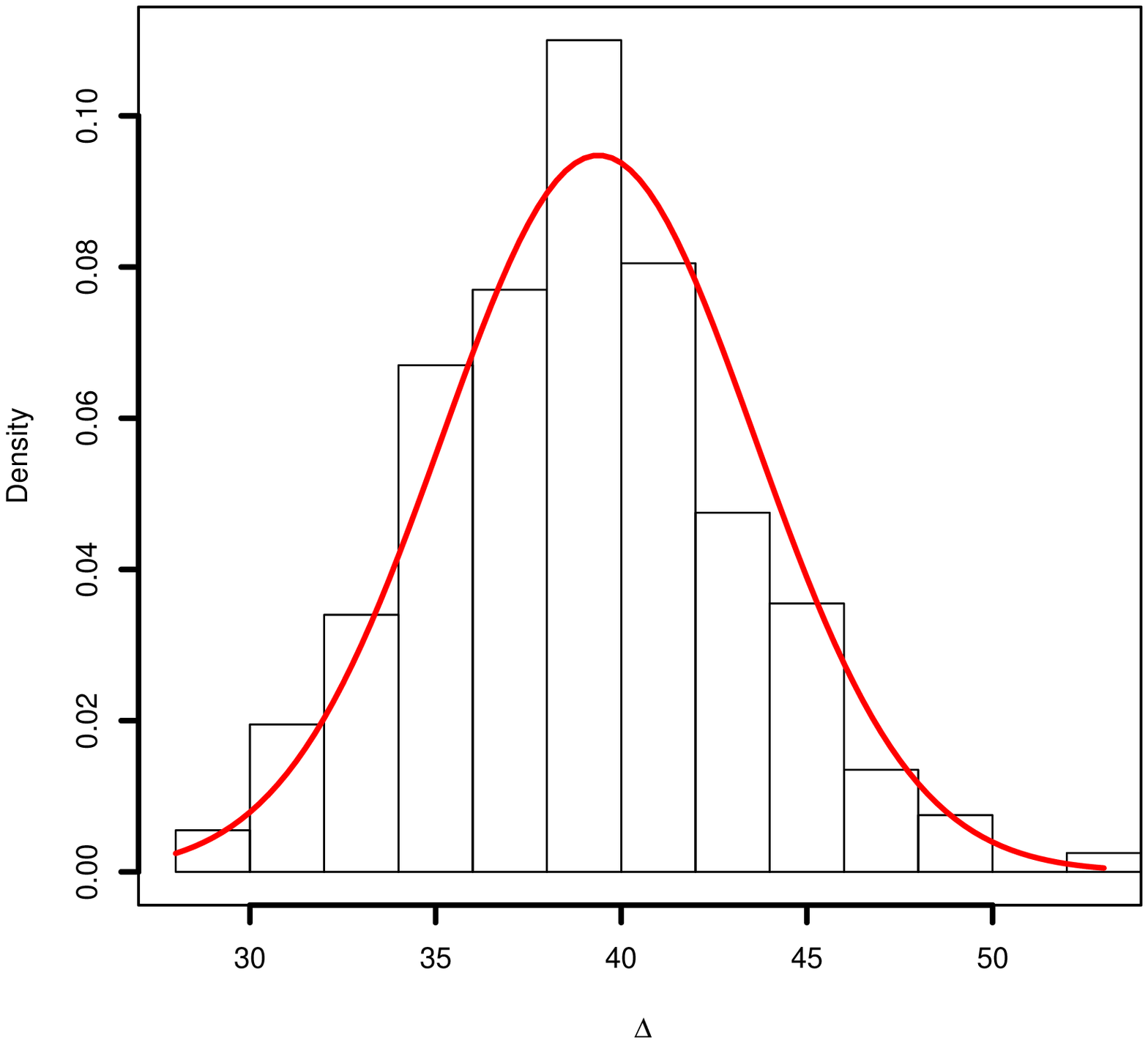}}
\caption{Distribution of the asymmetry $\Delta$ for the cases $M_{DM}=0.6\times 10^{12}$ M$_{\odot}$ (top panel), $0.7\times 10^{12}$ M$_{\odot}$ (central panel), $0.8\times 10^{12}$ M$_{\odot}$ (bottom panel), when the travel time is fixed to $330$ Myr.}
\end{figure}

HVSs should present an asymmetric distribution. Although the clear cutoff in the distribution can be contaminated by the hot BHB stars of the sample, a statistical asymmetry between the number of ingoing and outgoing stars is still expected \citep{per09}. We can search for the asymmetry $\Delta$ in the MMT stars sample by counting the number of stars beyond $\nu(r)$ and below $-\nu(r)$ for the different $M_{DM}$ and $t_*$. We define $\Delta_+$ as the number of outgoing stars that lie beyond the critical line $\nu(r)$ plus the stars with positive velocities that have Galactocentric distances larger than the zero-velocity point. On the other hand, $\Delta_-$ is defined as the number of ingoing stars that lie below the critical line $-\nu(r)$ plus the stars with negative velocities that have Galactocentric distances larger than the zero-velocity point. The asymmetry $\Delta$ is defined as $\Delta=\Delta_+-\Delta_-$.

As discussed, outside of the critical lines an asymmetry is expected because of the stars' finite lifetime and the MW potential. We vary $M_{DM}$ in the range of $[0.6$-$1.8]\times 10^{12}$ M$_{\odot}$ and $t_{*}$ in the range of $[300$-$640]$ Myr. For a fixed halo mass, we compute $\Delta$ for each value of $t_*$. In our calculations, we assume that all the stars have a maximal travel time $t_*$. This corresponds to the maximum travel time from the GC and determines, along with $M_{DM}$, the shape of the critical lines. In order to decide if a star lies below or beyond such lines, we compute error bars for distances and velocities from the data and generate random realizations inside the error bars. We take into account the fact that distances of MMT sample stars have large uncertainties, since they depend on the assumed absolute magnitude and stellar type. For the cases under analysis, we perform $1000$ Monte Carlo realizations of $\Delta$ to propagate the uncertainties, and fit the resulting distribution with a normal function. Then, we associate the mean of the distribution with $\Delta$ and the standard deviation with the uncertainty in $\Delta$. Figure 2 illustrates the distribution of the asymmetry $\Delta$ for the cases $M_{DM}=0.6$-$0.7$-$0.8\times 10^{12}$ M$_{\odot}$, when the travel time is fixed to $330$ Myr (see also Table 2). Figure 3 shows the resulting $\Delta$ for different halo masses as a function of $t_*$.

\begin{table}
\caption{Values of the asymmetry $\Delta$ in the case $t_*=330$ Myr}
\centering
\begin{tabular}{c c c}
\hline\hline
$M_{DM} (10^{12}$ M$_{\odot})$ & $\Delta$ & $\sigma_{\Delta}$ \\
\hline
0.6  & 31.28 & 4.89 \\
0.7  & 37.30 & 4.20 \\
0.8  & 39.38 & 4.21 \\
0.9  & 43.42 & 4.36 \\
1.0  & 44.60 & 4.07 \\
1.2  & 49.33 & 4.58 \\
1.4  & 53.31 & 4.65 \\
1.6  & 56.55 & 4.45 \\
1.8  & 60.57 & 3.98 \\
\hline
\end{tabular}
\end{table}

\begin{figure}
\centering
\subfigure{\includegraphics[scale=0.68]{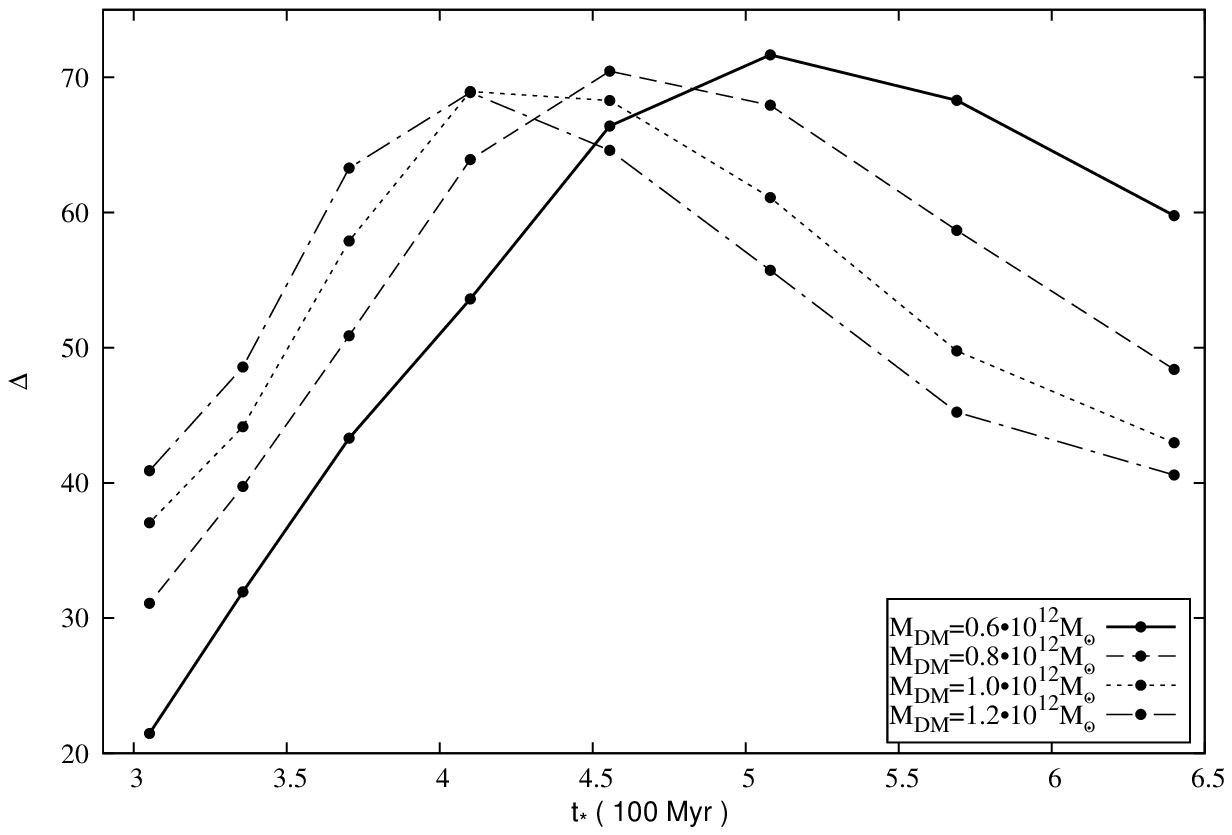}}
\subfigure{\includegraphics[scale=0.68]{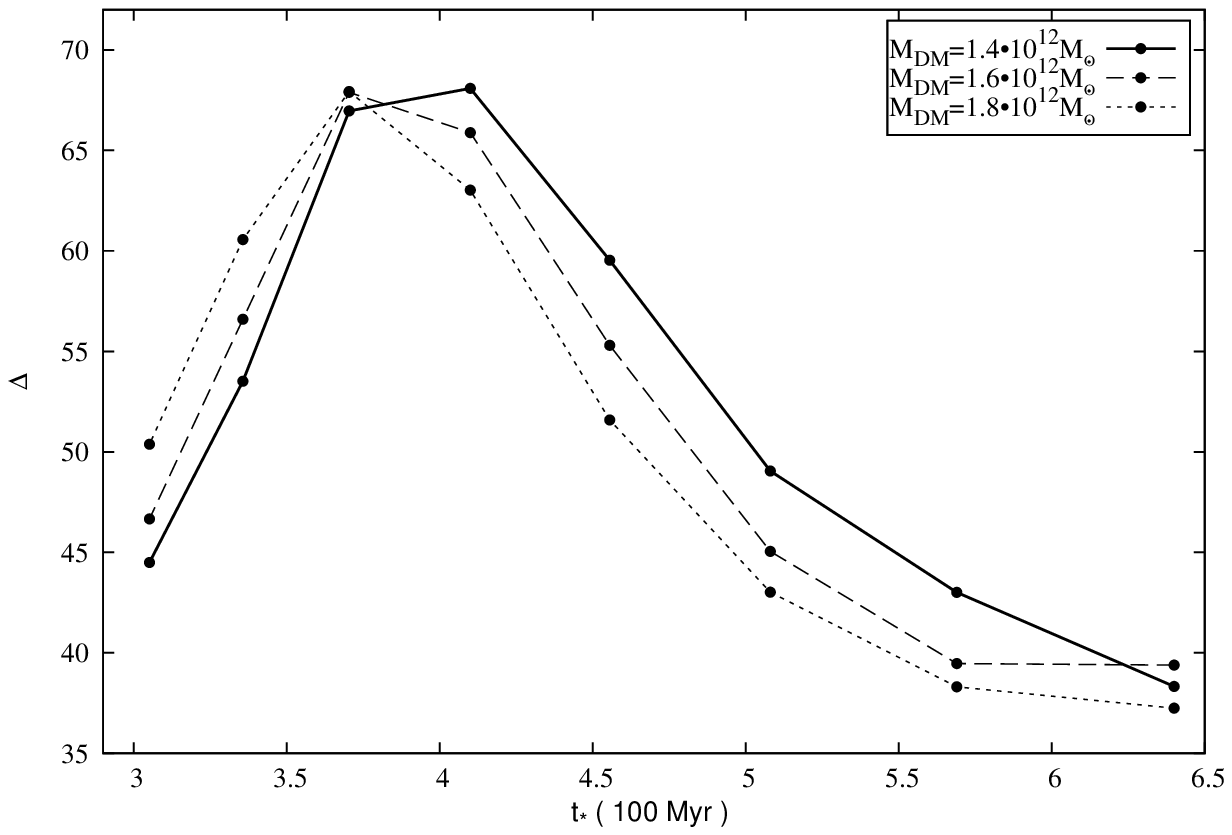}}
\caption{{$\Delta$} for different halo masses as a function of $t_*$, for $0.6 \le (M_{DM}/10^{12} \mathrm{M}_{\odot}) \le 1.2$ (top panel) and $1.4 \le (M_{DM}/10^{12} \mathrm{M}_{\odot}) \le 1.8$ (bottom panel).}
\end{figure}

\citet{brw07a} found a significant excess of stars travelling with $v_r>275$ km s$^{-1}$. They compared the asymmetry in the velocity distribution of the halo stars to the theoretical predictions of \citet{brm06}, and concluded that HVSs ejected both on unbound and bound orbits are the most plausible explanation for the observed excess of positive-velocity outliers. Bound HVSs are stars ejected from the Galactic center on bound orbits by the same mechanism that produces unbound stars, i.e. their ejection velocity does not exceed the local escape speed. Moreover, \citet{brw07a,brw07b} showed that, while stars with velocities $|v_r|<275$ km s$^{-1}$ are well described by a Gaussian distribution, the asymmetry of stars with $v>275$ km s$^{-1}$ is significant at the $5\sigma$ level. The choice of $275$ km s$^{-1}$ as threshold is motivated by the relative absence of stars with velocities less than $275$ km s$^{-1}$ in the MMT survey \citep{brw07a,brw07b,brw14}. The lack of stars moving at $v_r<-275$ km s$^{-1}$ suggests that the bound positive-velocity outliers in the sample have lifetimes less than the orbital turn-around time. \citet{brw14} presented $37$ stars that cause the excess of positive-velocity outliers (stars with $v>275$ km s$^{-1}$ in Table 1). Some of these outliers are unbound HVSs. By definition, a star is considered as an unbound HVS if its Galactocentric velocity exceeds the local escape velocity at its Galactocentric distance. Table 3 shows the number of HVSs, $\eta$, computed for different halo masses. Here $\eta$ depends on the assumed Galactic potential. Since $v_{esc}\propto M_{DM}^{1/2}$, different halo masses give different number of stars beyond the local escape velocity, with more massive haloes predicting a lower number of HVSs. Given $\eta$ for $M_{DM}=10^{12}$ M$_\odot$, the \citet{brw14} data suggests that the number of unbound $2.5-4$ M$_{\odot}$ HVSs is $\approx 300$ over the entire sky within $r < 100$ kpc ejected with a rate of $1.5\times 10^{-6}$ yr$^{-1}$. Figure 4 shows the escape speed curves for different $M_{DM}$ along with the MMT positive-velocity outliers with $v>275$ km s$^{-1}$. As in the case of $\Delta$, we take into account the uncertainties in stellar distances. In order to decide if a star lies below or beyond $v_{esc}$, we compute error bars for distances and velocities from the data and generate random realizations inside the error bars. We perform $1000$ Monte Carlo realizations of $\eta$ to propagate the uncertainties, and fit the resulting distribution with a normal function. Then, we associate the mean of the distribution with $\eta$ and the standard deviation with the uncertainty in $\eta$.

\begin{table}
\caption{Number of unbound ($\eta$) and bound ($\Gamma-\eta$) HVSs}
\centering
\begin{tabular}{c c c c}
\hline\hline
$M_{DM} (10^{12}$ M$_{\odot})$ & $\eta$ & $\sigma_{\eta}$ & $\Gamma-\eta$ \\
\hline
0.6  & 24.40 & 1.04 & 12.60 \\
0.8  & 21.97 & 0.81 & 15.03 \\
1.0  & 20.20 & 0.85 & 16.80 \\
1.2  & 18.81 & 0.60 & 18.19 \\
1.4  & 17.90 & 0.75 & 19.10 \\
1.6  & 16.56 & 0.83 & 20.44 \\
1.8  & 15.28 & 0.84 & 21.72 \\
\hline
\end{tabular}
\end{table}

As discussed above, \citet{brw14} presented $\Gamma=37$ stars that cause the excess of positive-velocity outliers at $v_r>275$ km s$^{-1}$, whose plausible explanation is outgoing bound and unbound HVSs \citep{brw07a,brw07b}. The asymmetry in the velocity-distance distribution is due to outgoing unbound and bound HVSs outside the critical lines region as a consequence of the cutoff in the ingoing HVSs. However, the clear cutoff in the ingoing stars may not be observable since it would be smeared by the halo stars contaminating the sample, but a statistical asymmetry between the number of ingoing and outgoing stars in the sample is still expected \citep{per09}. Hence, the estimated value of the asymmetry $\Delta$ that quantifies the number of bound and unbound HVSs must be compatible with the number of outliers that give the excess in the velocity distribution $\Gamma$. \citet{per09} suggested that another source for the asymmetry in the v-r plane may be the runaways from the Galactic disk \citep{sil11,brw15}. However, the MMT survey targets the Galactic halo and covers high Galactic latitudes $|b|\gtrsim 30^{\circ}$, where the only source of contamination is hyper-runaways \citep{brw15}. Hyper-runaways are stars ejected from the Galactic disk, probably as a consequence of multi-body interactions or supernovae explosions \citep{irr10}, with Galactic rest-frame velocities of the order of the local escape speed. \citet{heb08} found that the Galactic rest-frame velocity of the massive B giant runaway HD 271791 was larger than the local escape speed. The ejection rate of hyper-runaways with speeds comparable with HVSs is $\approx 8\times 10^{-7}$ yr$^{-1}$ \citep{brw15}. On the other hand, the ejection rate of HVSs is $\approx 10^{-5}-10^{-4}$ yr$^{-1}$ \citep{yut03}. Since the ejection rate of hyper-runaways is $10-100$ times smaller than HVSs rate, we expect $\approx 1$ hyper-runaway each $\approx 10-100$ HVSs \citep{per12}. Since this source of error is $\ll \Gamma$, we conclude that a possible contamination by hyper-runaways does not affect significantly our results. In conclusion, the favored model is the one for which $\Gamma$ is comparable to $\Delta$, within its error bars.

\begin{figure}
\centering
\includegraphics[scale=0.68]{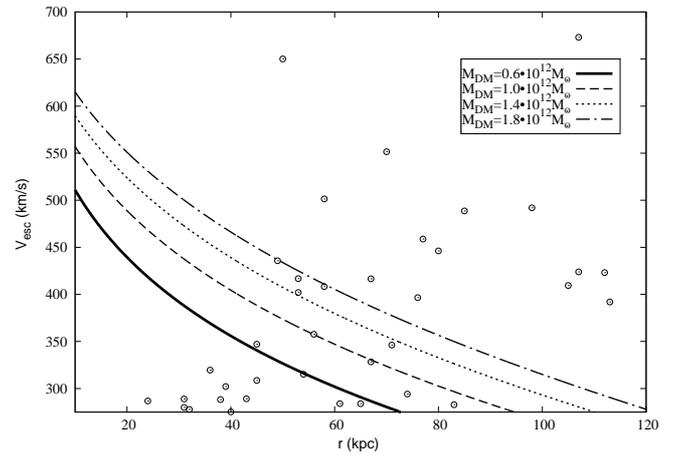}
\caption{Escape velocity curves for different halo masses. Circles are observed distances and velocities of the MMT star sample for stars with $v>275$ km s$^{-1}$ \citep{brw14}.}
\end{figure}

The asymmetry has a two dimensional dependence on $M_{DM}$ and $t_*$, which gives two possible interpretations depending on whether we fix the halo or the travel time. By fixing the halo mass we can estimate the preferred value of the travel time. As a consequence, since HVSs are probably MS stars and $m_*\propto t_*^{-\beta}$, we could evaluate the preferred mass of HVSs. On the other hand, by fixing the mass of the HVSs, we can constrain $M_{DM}$. Table 1 of \citet{brw14} presents also mass estimations for the unbound and bound populations of HVSs. The average mass of HVSs is $3.11$ M$_{\odot}$. If the travel time of HVSs is fixed to $330$ Myr (see Table 2), which corresponds to the average HVSs mass, the NFW halo mass parameter $M_{DM}$ is constrained to $(0.6$-$0.9)\times 10^{12}\ \mathrm{M}_{\odot}$, which gives a favored mass for the Milky Way in the range $(1.2$-$1.6)\times 10^{12}\ \mathrm{M}_\odot$ inside the virial radius $r_{200}$ \citep{kly99}. \citet{per09} suggest that an asymmetry of eight stars corresponds to the $1\sigma$ probability level. In this case, $M_{DM}$ is constrained to $(0.6$-$1.2)\times 10^{12} \mathrm{M}_{\odot}$, which yields a Milky Way mass in the range $(1.2$-$1.9)\times 10^{12} \mathrm{M}_\odot$.

\citet{gne10} used the MMT halo stars data (without including the outliers in the sample, i.e. the HVSs) to derive the Galactic circular velocity and the MW mass distribution. Adopting a three-component Galactic potential \citep{kly99}, they found that the \citet{brw10} data suggest a virial mass $M_{vir}=(1.6\pm 0.3)\times 10^{12} \mathrm{M}_\odot$ at the virial radius $r_{vir}=300$ kpc. Our inferred MW mass range is consistent with the \citet{gne10} results and recent independent determinations \citep*{mcm11,ead15,mcm16}.

Our results can be improved with better data on the mass and 3D velocity of HVSs. The tracers we studied suffer from the lack of tangential velocity measurements since the MMT survey is a spectroscopic survey. However, the European mission \textit{Gaia} (http://www.cosmos.esa.int/web/gaia) will be able to provide data also for proper motions, and hence tangential velocities, for some of the stars we considered.

\section{Conclusions}
\label{sec:con}

We have used the kinematics of HVSs as a method to constrain the MW mass. We studied the kinematics of HVSs observed by the MMT survey \citep{brw10,brw14} in the Galactic halo. The asymmetric velocity-distance distribution of the observed stars depends both on their lifetimes and the MW potential. We have found that, if the travel time of HVSs is fixed to $330$ Myr, which corresponds to the average HVSs mass, the halo mass parameter is constrained to the range $(0.6$-$1.2)\times 10^{12} \mathrm{M}_\odot$, which gives a favored mass for the Milky Way in the range $(1.2$-$1.9)\times 10^{12} \mathrm{M}_\odot$ inside the virial radius $r_{200}$.

\begin{acknowledgements}
We thank Warren Brown for supplying the data on the kinematics of the stars in the HVSs sample and for helpful comments on the paper. We also thank Scott Kenyon and an anonymous referee for insightful comments. GF acknowledges hospitality from the Institute for Theory and Computation at the Harvard-Smithsonian Center for Astrophysics, where the early plan for this work was conceived, and also thanks David Merritt for useful discussions and comments. GF acknowledges Sapienza University of Rome for the grant "Progetti Avvio alla Ricerca" (Starting Research Grant, 0051276), which funded part of this work.
\end{acknowledgements}

\end{document}